\documentclass[sigconf]{acmart}

\usepackage{multirow}
\usepackage{tcolorbox}
\usepackage{upquote}
\usepackage{xcolor}
\usepackage{pifont}
\usepackage{makecell}

\definecolor{good}{RGB}{0, 96, 0}
\definecolor{bad}{RGB}{96, 0, 0}

\newcommand{\desc}{\textit{desc}}
\newcommand{\apiseq}{\textit{apiseq}}
\newcommand{\qtt}[1]{``\texttt{#1}''}

\newlength{\twidth}

\newcommand{\good}{\textcolor{good}{\ding{51}}}
\newcommand{\bad}{\textcolor{bad}{\ding{55}}}
\newcommand{\goodrnn}{\good{} \textbf{RNN:} }
\newcommand{\badrnn}{\bad{} \textbf{RNN:} }
\newcommand{\goodcb}{\good{} \textbf{CodeBERT:} }
\newcommand{\badcb}{\bad{} \textbf{CodeBERT:} }
\newcommand{\actual}{\,\,\, \textbf{Actual:} }

\copyrightyear{2022}
\acmYear{2022}
\setcopyright{acmcopyright}\acmConference[ICPC '22]{30th International Conference on Program Comprehension}{May 16--17, 2022}{Virtual Event, USA}
\acmBooktitle{30th International Conference on Program Comprehension (ICPC '22), May 16--17, 2022, Virtual Event, USA}
\acmPrice{15.00}
\acmDOI{10.1145/3524610.3527872}
\acmISBN{978-1-4503-9298-3/22/05}

\AtBeginDocument{%
  \providecommand\BibTeX{{%
    \normalfont B\kern-0.5em{\scshape i\kern-0.25em b}\kern-0.8em\TeX}}}

\begin{document}

\title{Deep API Learning Revisited}

\author{James Martin}
\email{james.martin3@mail.mcgill.ca}
\affiliation{%
  \institution{McGill University}
  \city{Montréal}
  \country{Canada}
}
\author{Jin L.C. Guo}
\email{jguo@cs.mcgill.ca}
\affiliation{%
  \institution{McGill University}
  \city{Montréal}
  \country{Canada}
}


\begin{abstract}
  Understanding the correct API usage sequences is one of the most important tasks for programmers when they work with unfamiliar libraries. However, programmers often encounter obstacles to finding the appropriate information due to either poor quality of API documentation or ineffective query-based searching strategy. To help solve this issue, researchers have proposed various methods to suggest the sequence of APIs given natural language queries representing the information needs from programmers. Among such efforts, Gu et al. adopted a deep learning method, in particular an RNN Encoder-Decoder architecture, to perform this task and obtained promising results on common APIs in Java.  In this work, we aim to reproduce their results and apply the same methods for APIs in Python. Additionally, we compare the performance with a more recent Transformer-based method, i.e., CodeBERT, for the same task. Our experiment reveals a clear drop in performance measures when careful data cleaning is performed. Owing to the pretraining from a large number of source code files and effective encoding technique,  CodeBERT outperforms the method by Gu et al., to a large extent.
\end{abstract}



\maketitle

\section{Introduction}

When working with unfamiliar APIs, programmers frequently resort to learning resources and code examples to understand API usage sequences~\cite{xia2017developers}. Programmers normally start with forming queries based on their information need, e.g., ``JSON serialize an object'', and then search the API documentation or websites such as StackOverflow~\cite{stackOverflow} to identify answers to their queries or to similar questions.  This process can be long and ineffective when the queries represent more complex API usage, normally involving a sequence of API calls from the target library, either due to the incompleteness of the documentation, or the mismatch between the search queries and the descriptions or discussions of the relevant APIs. 

\begin{figure}[t]
\centering
\begin{tcolorbox}
\begin{verbatim}
import sys
from os.path import dirname, join as join_path

def sys_path():
  """ Add `./third_party` to `sys.path`.
  """
    
  dir = join_path(dirname(__file__), 'third_party')
  if not dir in sys.path:
    sys.path.insert(1, dir)
\end{verbatim}
\end{tcolorbox}
{\huge \big\Downarrow}
\begin{tcolorbox}
\textbf{Description:} \,\texttt{Add \textasciigrave ./third\_party\textasciigrave{} to \textasciigrave sys.path\textasciigrave.}
\textbf{API sequence:} \,\texttt{os.path.dirname}, \texttt{os.path.join}, \texttt{sys.path.insert}\par
\end{tcolorbox}
\caption[]{An example of a description and API sequence pair extracted from a Python function\protect\footnotemark. The description is considered as a proxy of programmers' query while the API sequence is the expected output for this learning task.\label{extraction_example_python}}
\end{figure}
\footnotetext{An actual example from https://github.com/00/wikihouse/blob/master/patch.py}

Meanwhile, numerous previous work has proposed to use learning-based approaches to support programmers for various tasks. Considering that code exhibits repetitive patterns such as the use of identifier names and idioms, and even the appearance of bugs~\cite{naturalness, allamanis2018survey, allamanis2014mining}, learning-based approaches have been adopted to support learning, writing, and fixing programs~\cite{gu, feng2020codebert, goues2019automated}. In particular, a ``deep API learning'' method proposed by \citet{gu} represents a recent major progress on the task of suggesting API sequences based on a given query and has been broadly cited since its publishing. Adopting an RNN encoder-decoder architecture, the deep API learning method maps natural language queries to sequences of Java Platform API calls to help programmers implement certain functions in Java. Considering the demonstrated performance and substantial impact of this work, we adapt this technique to APIs from common libraries in Python, one of the most used programming languages. Starting from Python source code, we can use a similar technique to formulate the API sequence learning task as shown in Figure~\ref{extraction_example_python}. As a reproducibility study, our work investigates if a similar performance in terms of BLEU scores can be obtained for our curated Python dataset. Our experiment reveals that the performance on Java and Python datasets is comparable if a duplication removal step is performed on both datasets.

Furthermore, since the deep API learning work was published, more advanced techniques have been employed for code and natural language representation learning. For example, the work CodeBERT proposed by \citet{feng2020codebert}, using a transformer-based model and a pretraining phase with a large number of code-natural language pairs, demonstrated an improved performance on several downstream tasks that require reasoning across natural language and programming languages. Its potential on the task of API sequence generation, however, is yet unknown. Therefore, we also include CodeBERT in our investigation to compare with the original deep API learning approach. We aim to understand if API sequence learning can also benefit from a transformer-based architecture and pretraining tasks on a large dataset. We observe that CodeBERT indeed performs better by a large margin on both the Python and Java datasets. The improvement over the RNN based model is still substantial for Python even without the pretraining step.

In summary, we make the following contributions:
\begin{enumerate}
    \item Our experiment results have validated the reproducibility of the work by \citet{gu} for the task of API learning on the original Java dataset and our newly curated Python dataset; 
    \item Our study reveals the importance of data inspection and how the quality of the dataset might impact model performance and its interpretation;
    \item The comparison with the CodeBERT model shows a clear potential for adopting transformer-based models for the API learning task.
    \item The Python dataset we have curated is shared as supplementary material for future research in this direction.\footnote{Datasets are available at \url{https://doi.org/10.5281/zenodo.6388030}.  The code that was used to curate the dataset is available at \url{https://github.com/hapsby/deepAPIRevisited}.}
\end{enumerate}




\section{Related Work}
\subsection{Machine Learning for Code}
Since the early work revealing the naturalness of source code~\cite{naturalness}, numerous efforts have been proposed to accumulate and learn from large datasets of source code and peripheral artifacts for various programming related tasks. The tasks normally either focus on the code itself, such as code completion~\cite{svyatkovskiy2021fast}, bug detection~\cite{li2019improving}, and program repair~\cite{yasunaga2020graph}, or focus on the relationship between code and natural language and code written in a different programming language, such as commit message generation~\cite{jiang2017automatically}, tracing source code with issues~\cite{rath2018traceability}, code summarization~\cite{wan2018improving}. We suggest that interested readers refer to surveys by \citet{allamanis2018survey} and by \citet{devanbu2020deep} to get a more comprehensive picture. Among such a variety of machine learning for code applications, in particular adopting deep learning based approaches, we focus on connecting natural language descriptions of code functions to a subset of source code, in particular, API calls, an idea first proposed by \citet{gu}. 

Recent work on machine learning for code has a clear emphasis on some sort of self-supervision to pretrain the models. For example, the work of CodeBERT was pretrained using two self-supervised objectives across natural language and source code~\cite{feng2020codebert}. \citet{gu2022assemble} propose to assemble CodeBERT and other similar ``foundation'' models to perform the code summarization task. \citet{allamanis2021self} performs the self-supervision on bug detection and repair by training two models at the same time, i.e., a selector model that injects bugs in the code and a detector model that detects and fixes the bug. In this work, we adopt the pretrained CodeBERT model for the API learning task. We also explicitly investigate the impact of the pretraining step on the rate at which the CodeBERT model converges when trained with the Python API dataset.

\subsection{Programmers' API Learning Support}
Previous work on supporting programmers to learn API usage can be categorized into three groups. One line of work focuses on augmenting existing API documentation from external resources such as StackOverflow~\cite{treude2016augmenting}. Since the official documentation normally suffers from quality problems such as incompleteness, out-of-datedness, and correctness~\cite{aghajani2019software}, external resources can bring additional insights to the programmers for the target API.  The second direction is on improving code example search. By comparing the representation of search queries with code snippets, the objective of such work is to recommend the most relevant code examples~\cite{bai2020graduate}. The third direction is based on generative machine learning methods using which the search queries from the programmers can be ``translated'' into source code or API sequences. The work by \citet{gu} upon which our study is built can be put in the last direction. While the current machine learning models such as CodeBERT might be able to generate the entire code sequence, we select the task with a focus on API sequence considering its clear implication on programmers' learning objective.

\section{Experiment Design}
\subsection{Data Collection}
The data set for training and evaluating learning based models consists of (\desc, \apiseq) pairs, where \desc{} is a natural language description of a task, representing the queries that programmers might use, and \apiseq{} is a sequence of calls to API or library functions.  While there are existing datasets of code and natural language pairs such as the CodeSearchNet~\cite{husain2019codesearchnet}, there is no existing dataset for Python that connects the natural language description to isolated API calls, to the best of our knowledge. Obtaining such a dataset is non-trivial and requires the resolution of import aliases; therefore we curated from GitHub ourselves. We first downloaded 257,049 open source Python projects from GitHub and extracted each top-level function and class method from these projects. We then converted each to a (\desc, \apiseq) pair. We discuss the detailed process below.  

\subsubsection{Projects Selection}

We started with identifying all projects tagged as Python projects on GitHub. We then ranked them based on stars and selected only the projects that have at least five stars indicating obtaining certain popularity. We then removed all the forks from other projects to avoid duplicated projects. We also filtered projects with a size bigger than 300 MB due to limited disk space and bandwidth. For the 257,049 projects that remained after filtering, we cloned the latest revision of the default branch without submodules for further processing in the next step.  GitHub did not offer a way to distinguish between Python 2 and Python 3 projects, so projects written in both versions of Python were collected.  However, if a particular source file did not contain valid Python 3 syntax, the source file was rejected in a later step.

In the original deep API learning experiment, projects were downloaded with at least one star; we chose to require more stars, because the number of Python projects available on GitHub greatly exceeded the number of projects that we were able to download in our timeframe.

\subsubsection{Descriptions Extraction}

We used the docstring for a Python function, if it exists, to extract its description. Each description is the concatenation of the ``primary description'' and the ``returns description'' -- The ``primary description'' is the beginning content of the docstring until either an empty line or a line starting with \qtt{param}, \qtt{params}, \qtt{parameter}, \qtt{parameters}, \qtt{return}, or \qtt{returns} (with possible decorating colons); The ``returns description'' is the part of the docstring the first line of which begins with \qtt{return} or \qtt{returns} (with possible decorating colons) until a blank line or a line beginning with \qtt{param}, \qtt{params}, \qtt{parameter}, or \qtt{parameters}.  In the latter case, the decorating colons are removed and the word is normalized to \qtt{return}.

When the docstring is not available, we extracted the descriptions from the function name. Underscores in the function name were turned to spaces, and spaces are inserted before uppercase letters, which are then lowercased. For example, the function names \qtt{do\_something} and \qtt{doSomething} both become \qtt{do something}.

The extraction was performed using the \texttt{ast} module in Python 3.8.8 to parse every file with a \texttt{.py} filename extension found in each project, in total 9,292,645 such files. During this process, we removed 345,253 files due to character encoding errors, and 610,892 files due to contained syntax errors, including errors on Python~2 syntax that cannot be recognized by Python~3's \texttt{ast} module. 

In the original deep API learning experiment, the description of a function was extracted from the first sentence of the function's JavaDoc comment, and functions with no JavaDoc were ignored.  Our extraction procedure differs from the original to compensate for our smaller corpus of projects (257,049 projects instead of 442,928 projects) and the relative sparsity of docstrings in Python code as compared to JavaDoc comments in Java code.


\subsubsection{API Sequence Extraction}
\label{importstatements}

To extract the API call sequence from the function body, we first processed the import statement in each python source code file. 
When an import statement is encountered, we recorded the symbols defined by the import statement along with the appropriate replacement string, unless:
\begin{itemize}
\item it is a relative import, e.g., \qtt{from ..x import y};
\item  the module being imported is part of the project's source code, e.g., \qtt{import a.b} when either the file \qtt{a/b.py} or the file \qtt{a/b/\_\_init\_\_.py} exists; or
\item  it is a wildcard import, e.g., \qtt{from x import *}.
\end{itemize}
In the first two cases, the import is ignored because calls to functions in these modules are not considered to be API calls.  We also ignored the last case considering the complexity of matching API calls; it could be added in future versions.

Once the module name is extracted, we searched the body of each function for all function calls to the imported modules. Each function call consists of one or more identifiers joined by dots, followed by a list of arguments: for example, \texttt{x.y.z()}.  If the function call matches a top-level function that was defined earlier in the file or a local function that was defined earlier in the function, then the function call was replaced with the list of API calls extracted from that earlier function.  Otherwise, we checked whether the first identifier (e.g., the identifier \texttt{x} in the function call \texttt{x.y.z()}) was defined by any target import statements for the file.  If so, then the first identifier was replaced according to the import rule, and the function call is then considered as an API call.

We identified all the API calls for each function and listed them in the order in which the close-parentheses of their argument lists appear in the source code: for example, if the source code reads \texttt{f(g(),h())} then the function calls are listed as \texttt{g}, \texttt{h}, \texttt{f}. If no API calls were extracted from a function, we removed that function from further analysis. Among 68,048,224 functions, we extracted 28,493,702 instances of API sequences.

\subsubsection{Data Filtering}
\label{subsubsec:data_filtering}
From the initial dataset of (\desc, \apiseq) pairs, we further filter the instances to improve its quality and its applicability to the API learning task. In particular,
\begin{itemize}
\item we remove all pairs with more than 14 API calls, because the Deep API project truncates sequences to 28 tokens, and API calls each require at least two tokens. This leaves 27,296,964 pairs;
\item we sort all Python modules in descending order of popularity (determined by the total number of calls to functions in the module) and, in that order, we add each module to a list of accepted modules if the total number of identifiers in the API calls of this and all previously accepted modules does not exceed 9,995.  Then we remove all pairs that contain at least one API call to a module that is not in the list of accepted modules.  This guarantees that the output vocabulary, which consists of the identifiers together with the \qtt{.} separator and four other special tokens, does not exceed 10,000, which is the size of the RNN network's output vector and is therefore the maximum vocabulary size.  This step leaves just 7,505,321 pairs;
\item we remove all duplicate pairs, which leaves 1,655,645 pairs;
\item we remove all pairs with descriptions smaller than 3 words, which leaves 1,087,673 pairs; and
\item we remove all pairs with ``test'' in the function name or description.
\end{itemize}

The final dataset is comprised of 855,107 (\desc, \apiseq) pairs for Python programs. We set aside 4\% of the pairs to use as the test set and the remaining 820,967 pairs as the training set for training the deep learning models on the API learning task.

\subsection{Model Selection}

Two neural network models were selected in our experiment. We considered the RNN Encoder-Decoder model because it was proposed by \citet{gu} to perform API sequence learning for Java in their original work. We also investigated the CodeBERT model considering its broad impact and wide availability through GitHub~\footnote{\url{https://github.com/microsoft/CodeBERT}} and HuggingFace API~\footnote{\url{https://huggingface.co/microsoft/codebert-base}}. We describe how we trained and fine-tuned the models in detail below.

\subsubsection{RNN Encoder-Decoder}\label{rnn}

We have adapted the implementation of the Deep API project by \citet{gu} to the task of generating Python API sequences for given queries. This model has an RNN encoder that accepts embedded descriptions, a linear layer, a $\tanh$ layer, and an RNN decoder that outputs embedded API sequences.

The RNN encoder has an input size of 120 (the size of an embedded word) and one bidirectional hidden layer with 2000 features in total.  The linear layer reduces the 2000 features from the RNN encoder to 1000 features.  The RNN decoder has one hidden layer of 1000 features, an attention layer, and an output size of 10,000 which is our vocabulary size.

We trained this model in batches of 100 pairs. Porter's stemmer~\cite{Porter1980AnAF} was used to pre-process the input tokens before encoding. The AdamW algorithm \cite{adamw} was used to optimize the model parameters, with a learning rate $\alpha=10^{-4}$ and $\epsilon=10^{-8}$.  The loss function was the cross-entropy between the actual \apiseq{} and the \apiseq{} predicted by the model.  After every 1000 batches, the model was evaluated with the test data set: the actual \apiseq{}s from the test set were compared to the \apiseq{}s predicted by the model. We adopt the same metric as in the original deep API learning work by \citet{gu}, i.e., BLEU-4 score \cite{bleu} that is the BLEU score taking into account $n$-grams up to $n=4$.

\begin{figure}[t!]
  \centering
  \includegraphics[width=0.90\linewidth]{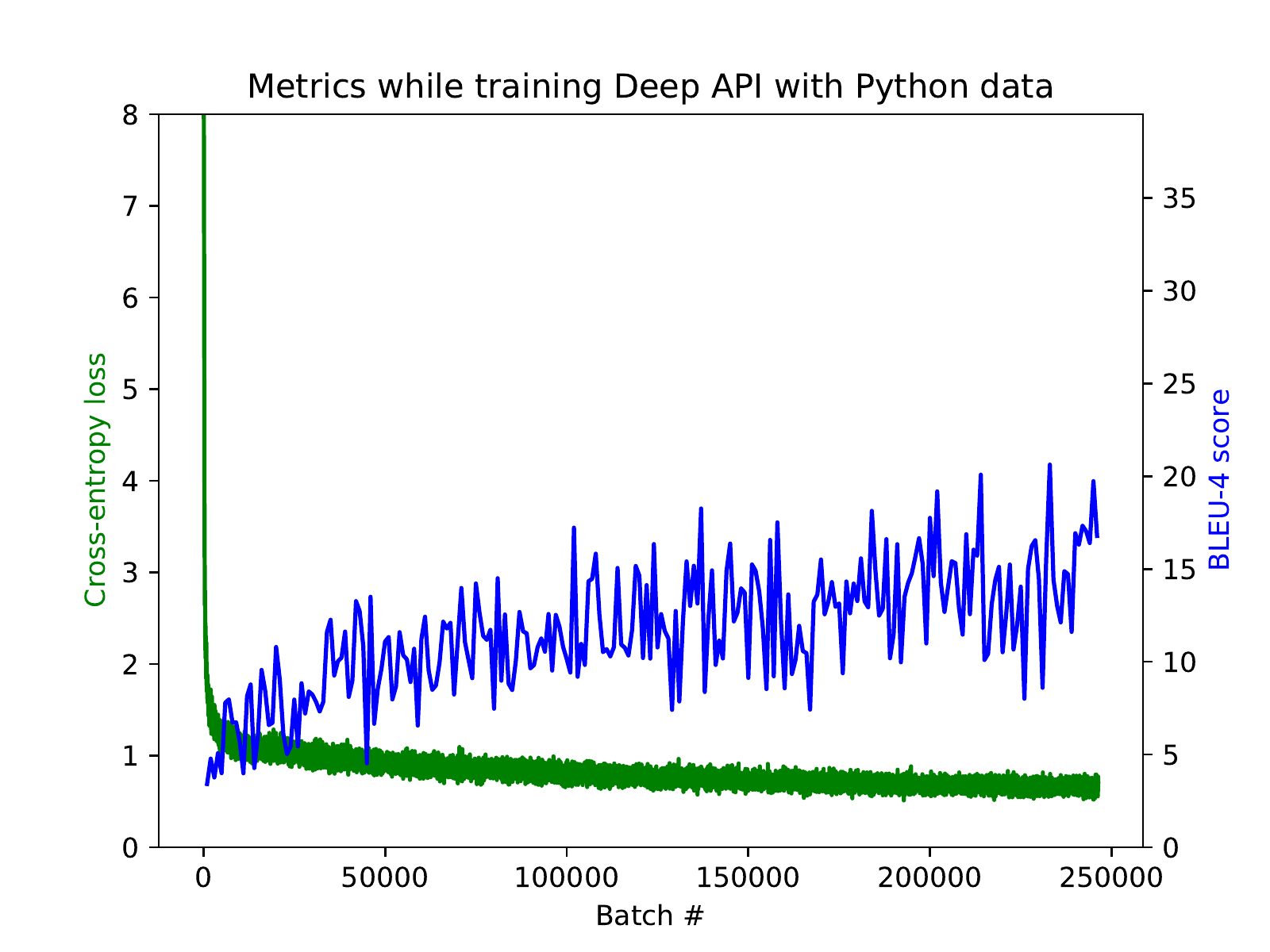}\hfill
  \caption{The cross-entropy loss and the BLEU-4 score achieved during the training of the RNN Encoder-Decoder with the Python API training set. Each BLEU-4 score is evaluated using a sample of 100 pairs from the testing set.\label{deepapi-graph_python}}
  \vspace{10pt}
  \includegraphics[width=0.90\linewidth]{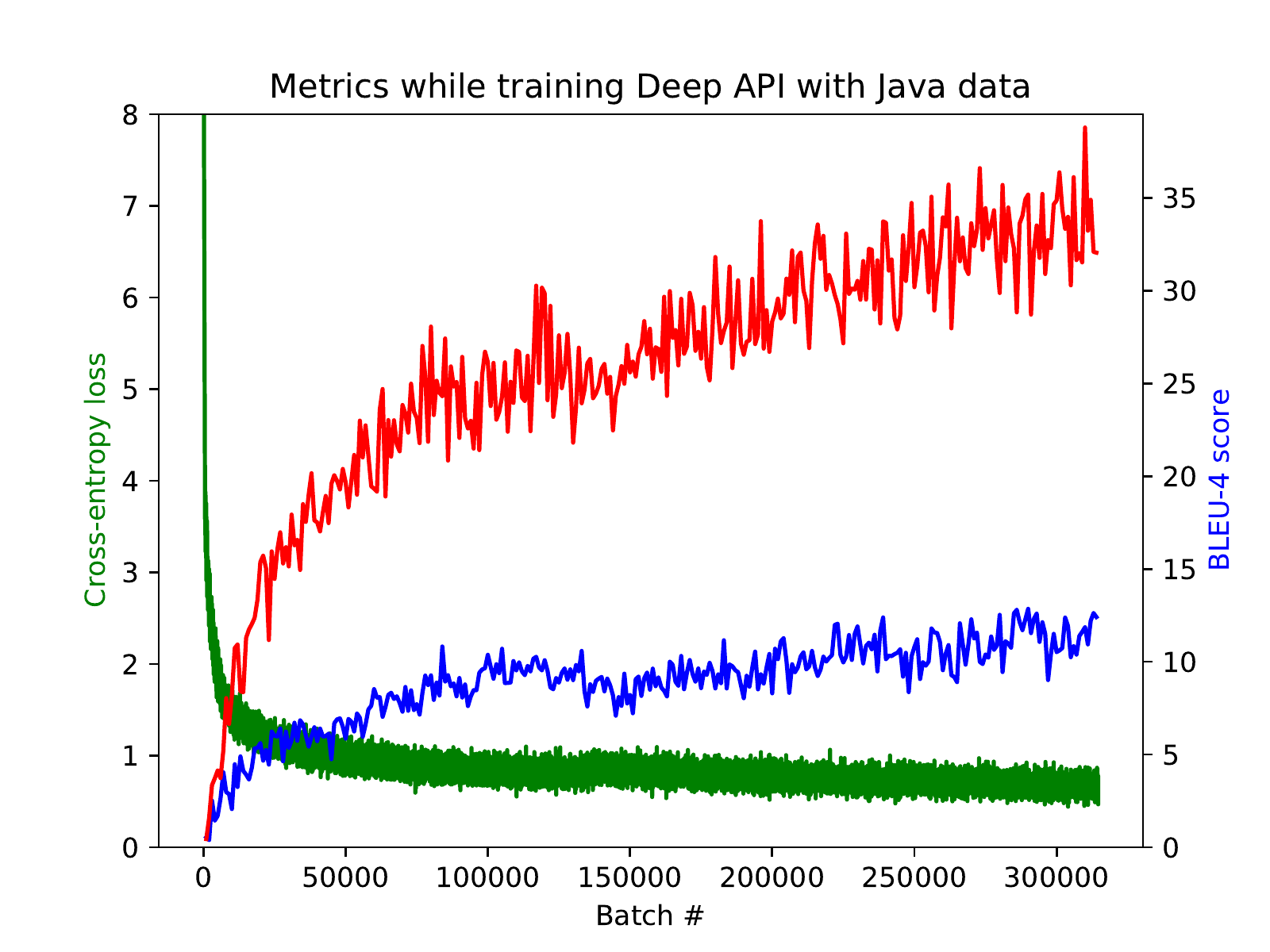}
  \caption{ The cross-entropy loss and the BLEU-4 score metrics achieved during the training of the same RNN Encoder-Decoder with the Java API training set provided by \citet{gu}\label{deepapi-graph_java}, both with and without duplicates removed. The BLEU-4 score achieved using the original test set is shown in red and using deduplicated test set in blue (a sample of 100 pairs in each case).}
\end{figure}

For performance reasons, only 100 test pairs were used for each evaluation of the BLEU-4 score shown in Figures \ref{deepapi-graph_python} and \ref{deepapi-graph_java}.  The final BLEU-4 scores in Table \ref{tab:all_bleu_scores} were computed with 10,000 pairs from the original test set, except for the deduplicated Java API test set which has only 2,441 pairs.  Each RNN model took approximately six days to train using an Intel\textsuperscript{\textregistered} Core™ i5-3570K (3.40GHz) CPU and a NVIDIA\textsuperscript{\textregistered} GeForce GTX™ 1070 GPU with 8 GiB VRAM.

\subsubsection{CodeBERT}\label{codebert}

We have adapted the code of the CodeBERT project \cite{feng2020codebert} to the task of generating Python and Java API sequences based on given queries. CodeBERT model reuses the model architecture of RoBERTa~\cite{roberta} and is pretrained with two objectives directly connecting the natural language and source code modalities:  
\begin{itemize}
\item A ``Masked Language Modeling (MLM)'' objective, where the model is given a natural language sequence and a code sequence with 15\% of the tokens replaced by a special \texttt{[MASK]} token, and the model infers what the original tokens were; and
\item a ``Replaced Token Detection (RTD)'' objective, where the model is given a natural language sequence and a code sequence with some of the tokens replaced by random tokens from the vocabulary, and the model determines which tokens were replaced.
\end{itemize}
The code sequences used to pretrain the model were from CodeSearchNet~\cite{husain2019codesearchnet}, a large dataset drawn from open source projects hosted on GitHub in six languages:  Go, Java, JavaScript, PHP, Python, and Ruby.  

We fine-tuned the pretrained model with our training set of Python (\desc, \apiseq) pairs.  The model was trained in batches of 96 pairs.  To ensure that the words in the task descriptions match the words that were used in the pretraining, we used the default RoBERTa tokenizer instead of the Porter stemmer.  The loss function was the cross-entropy function.  After every 1000 batches, the model was evaluated by computing the BLEU-4 score of our test set. 

We additionally experimented with CodeBERT model initialized with random parameters -- we directly trained the CodeBERT model with our Python (\desc, \apiseq) dataset. By comparing the model performance in both settings, we aim to understand the impact of the pretraining step with a larger dataset using self-supervised objectives.

10,000 pairs from the Python API test data set were used in the evaluation of each BLEU-4 score.  All 2,441 pairs from the deduplicated Java API test data set were used. The Python API CodeBERT model was trained for approximately 21 hours using an AMD\textsuperscript{\textregistered} EPYC™ (2.50 GHz) CPU and a NVIDIA\textsuperscript{\textregistered} RTX™ A6000 GPU with 48 GiB VRAM.  The Java API CodeBERT model was trained for approximately 60 hours on the same machine.

\section{Experiment Results}
\label{sec:results}
Our experiment is set up to answer the following research questions:

    - \textit{RQ1: To what extent can we reproduce the deep API learning results from \citet{gu} on the Java dataset?}

    - \textit{RQ2: Can the RNN Encoder-Decoder model achieve similar performance on the API learning task for Python compared with for Java?}
    
    - \textit{RQ3: To what extent can the CodeBERT model compete with the RNN Encoder-Decoder model on the API learning task?}
    
    \hspace{10pt} - \textit{RQ3.1:  To what extent does the CodeBERT model benefit from the self-supervised pretraining step}.

In this section, we discuss our findings on each research question in detail.  These results are summarized in Table~\ref{tab:all_bleu_scores}.

\begin{table}[t]
\caption{A summary of the top-1 BLEU-4 scores achieved with the various datasets and models tested.  The last column lists the outcomes when duplicates of the training pairs are not removed from the test set. \label{tab:all_bleu_scores}}
\includegraphics[width=0.90\linewidth]{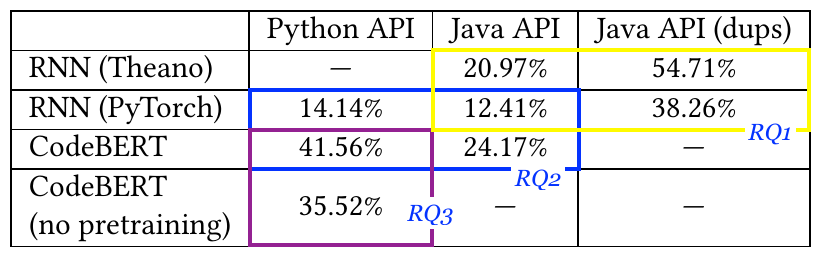}

\end{table}     

\subsection{Reproducing the Deep API Learning Results (RQ1)}
\label{subsec:eval_deepapi}

As a reproducibility study, our first step is to understand the extent to which we can reproduce the deep API learning results from \citet{gu} on the Java dataset used in the original work (RQ1).
The Deep API source code provided by \citet{gu} is available in two versions: an older implementation using Theano~\footnote{\url{https://github.com/Theano/}}, and a newer implementation using PyTorch~\footnote{\url{https://pytorch.org/}}.  The Theano implementation is exactly what was used in \citet{gu}.  We trained the Theano implementation with the data provided by \citet{gu} and achieved a top-1 BLEU-4 score of 54.71\%, consistent with that paper.

However, when we created the Python dataset, we noticed a large number of duplicated instances that cannot be removed on the project level (see Section~\ref{subsubsec:data_filtering}). This might be due to the reuse and clones of source code within and across projects. We then performed duplicate detection on the Java dataset released by the previous work and had a similar observation. The Java training set with 7,475,850 (\desc, \apiseq) pairs becomes 1,880,472 pairs after removing duplicates.  The Java testing set with 10,000 pairs becomes 2,441 after removing pairs that are also found in the training set.  The experiment was repeated, with one million iterations, and the BLEU-4 scores achieved were 20.97\% (top-1), 32.42\% (top-5), and 36.99\% (top-10).  This dramatic difference is due to the overlap between the training set and the test set in the previous work.  In our remaining experiment, we decide to compare with the BLEU-4 scores on the dataset after removing duplicates, as they more accurately reflect the model's capacity to generalize on unseen queries.

\subsection{RNN Encoder-Decoder Model Performance (RQ2)}
\label{subsec:rnn_result}
After establishing the baseline performance on the Java dataset, we used the PyTorch implementation developed by \citet{gu} to train the model with our Python data to answer our RQ2.  The BLEU-4 scores achieved are shown in Figure~\ref{deepapi-graph_python}. The model converged quickly and continued to improve with training with more batches. The BLEU-4 score achieved on the testing set was 14.14\% with the top-1 output API sequence. 

We also used the PyTorch implementation to train the original Java API dataset (with duplicates removed).  The results are shown in Figure~\ref{deepapi-graph_java}. The BLEU-4 score achieved was 12.41\%, which is less than the score of 20.97\% reported in Section~\ref{subsec:eval_deepapi}; possible reasons for this are discussed in Section \ref{discussion}.  

Figure~\ref{deepapi-graph_java} also shows the result of training the model with the original data set before the duplicates were removed.  The BLEU-4 score achieved was 38.26\%.

Our experiments indicate that the same model architecture performs similarly for Python and Java. At the same time, the evaluation of the model's performance is greatly affected by the quality of the test set. When it makes inferences on unseen queries, its performance is much weaker than on a collection of queries with entries from the training set.

\begin{figure}[t]
  \centering
  \includegraphics[width=0.90\linewidth]{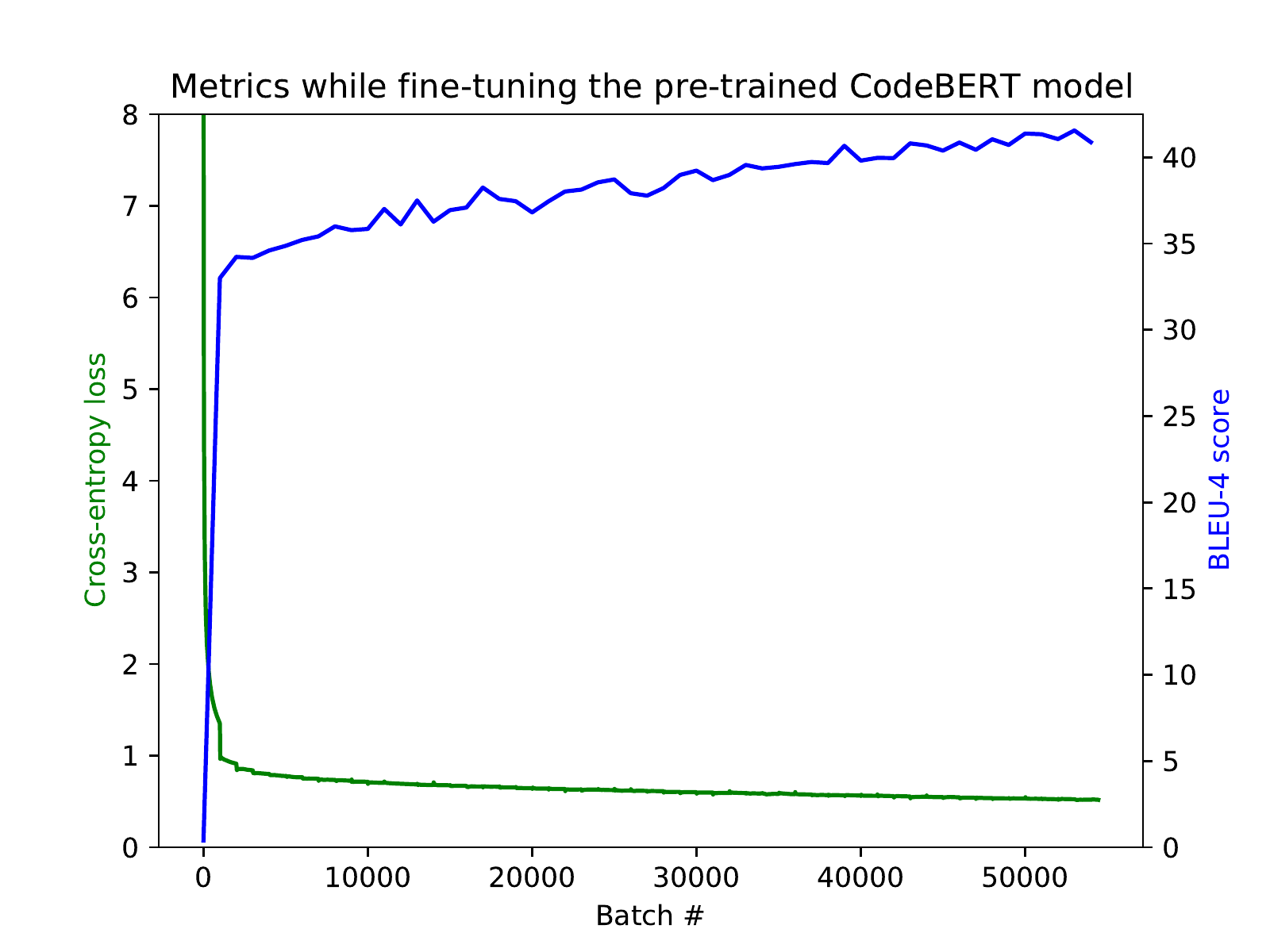}\hfill
  \caption{The cross-entropy loss and the BLEU-4 score achieved during the fine-tuning of the CodeBERT model with the Python API training set. \label{codebert-graph_python}}
  \vspace{10pt}
    \includegraphics[width=0.90\linewidth]{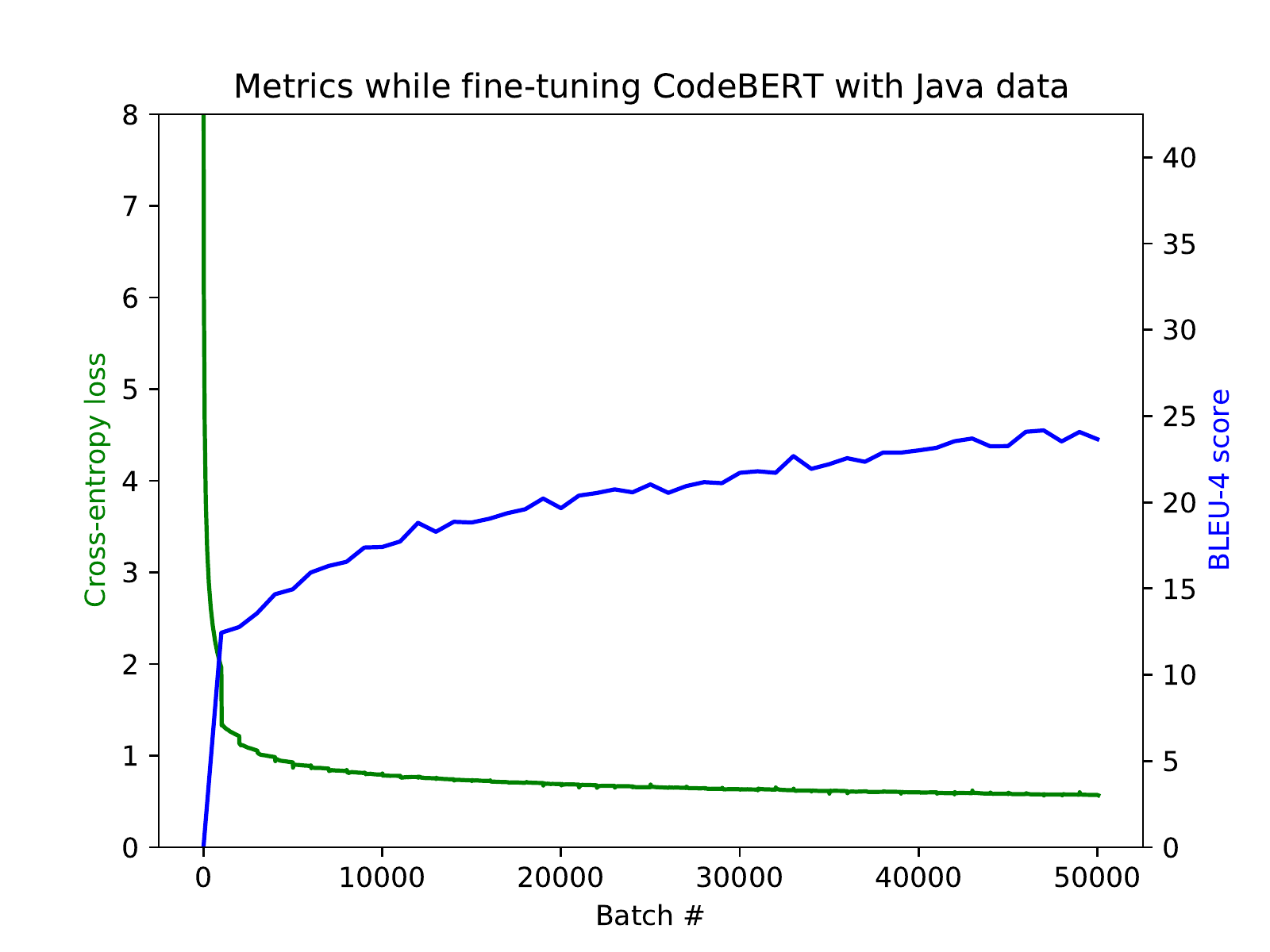}\hfill
  \caption{The cross-entropy loss and the BLEU-4 score achieved during the fine-tuning of the CodeBERT model with the Java API training set. \label{codebert-graph_java}}
\end{figure}

\begin{figure}[t]
  \centering
  \includegraphics[width=0.90\linewidth]{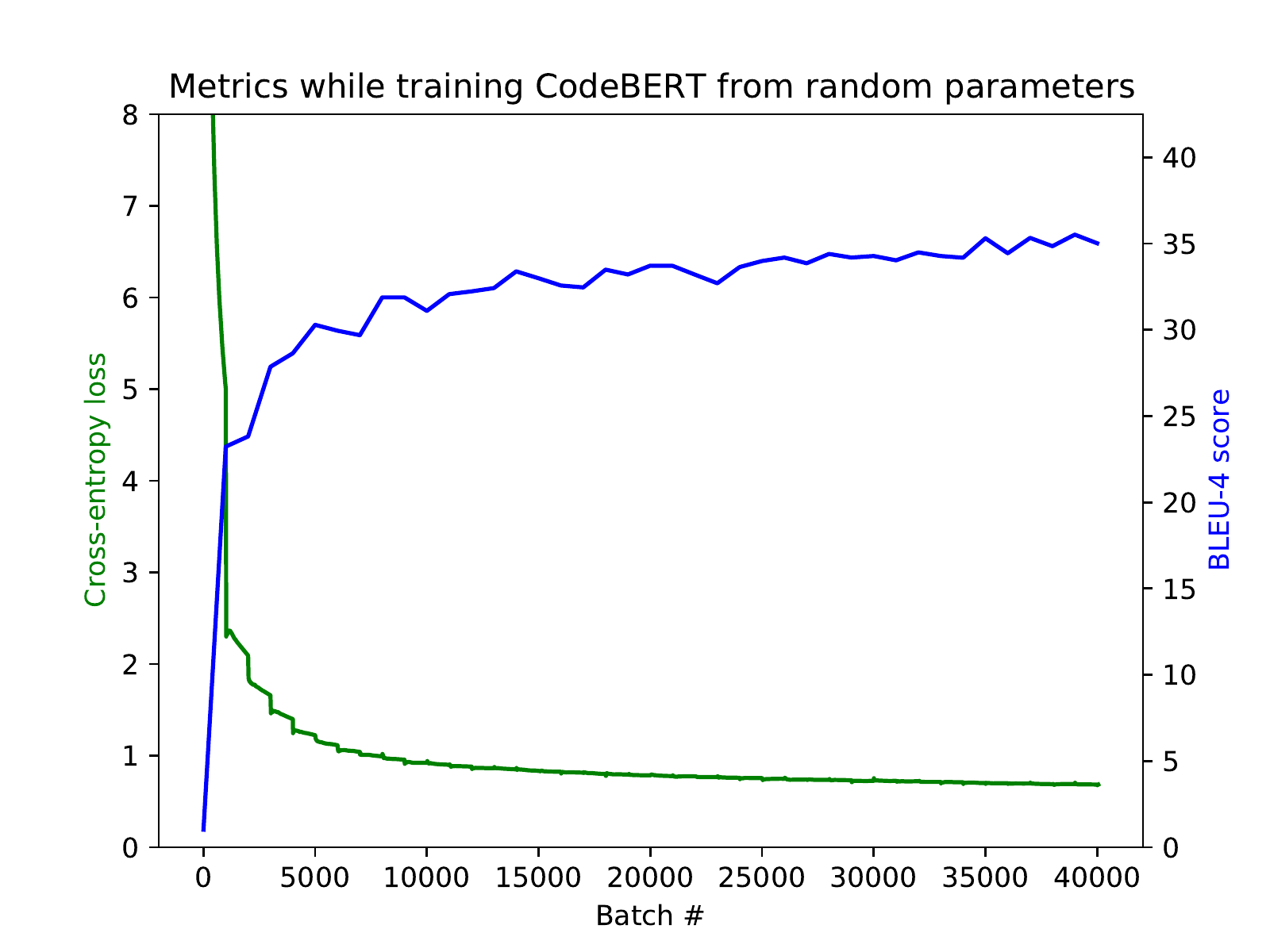}
  \caption{The cross-entropy loss and the BLEU-4 score achieved achieved when training the CodeBERT model with the Python API training set from random initial parameters (that is, discarding the pretraining). \label{codebert-graph_resetparams}}
\end{figure}

\subsection{CodeBERT Performance (RQ3)}
\label{subsec:codebert_result}
Our RQ3 aims to understand the potential of using a more recent approach, CodeBERT, on this task, compared to the method proposed by \citet{gu}. The performance of fine-tuning the pretrained CodeBERT model on the task of API sequence generation is shown in Figure~\ref{codebert-graph_python} and Figure~\ref{codebert-graph_java}. The maximum BLEU-4 score achieved on the Python dataset was 41.56\%, largely improved compared with 14.14\% achieved by the RNN Encoder-Decoder model. On the Java dataset, the maximum BLEU-4 score was 24.17\%. In comparison, the performance by the RNN Encoder-Decoder model on Java is only 12.41\% (using the Pytorch implementation). Such improvement might be attributed to the model architecture, i.e., transformer~\cite{transformer} versus RNN based models, and/or the pretraining with self-supervised objectives on a larger and potentially overlapped dataset. To understand the impact of each factor, we performed two additional analyses, one on the dataset overlapping and the other on model performance without pretraining. 

The CodeSearchNet dataset, on which the CodeBERT model is pretrained, was collected from GitHub, the same source as our Python dataset. It is possible that CodeSearchNet has seen a function during its pretraining step from which our test instance is constructed. To understand the overlap, we compare the \desc{} and \apiseq{} with CodeSearchNet. We found 93 \texttt{docstring}s in CodeSearchNet that exactly match a \desc{} in our Python test set after both were tokenized with the RoBERTa tokenizer.  We found 10,267 instances in CodeSearchNet in which the code matches an  \apiseq{} in our Python test set (after RoBERTa tokenization).  Matching for the \apiseq{} means that every token in the tokenized \apiseq{} is found in the tokenized CodeSearchNet code and is in the same order; however, the tokenized CodeSearchNet code may have additional tokens. When both \desc{} and \apiseq{} from one instance is matched to a CodeSearchNet instance, we consider this pair to be matched. We found only one matched pair for the Python dataset. The same analysis was also performed on the Java dataset and the overlap is summarized in Table~\ref{tab:codesearchnet_overlap}.

\begin{table}[t]
\caption{Overlap between the testing dataset used for the API learning task and the CodeSearchNet pretraining data. \label{tab:codesearchnet_overlap}}
\begin{tabular}{|l|c|c|c|}
\hline
\multirow{2}{*}{} & \multirow{2}{*}{\begin{tabular}[c]{@{}c@{}}\# of Matched  \\ \desc{} \end{tabular}} & \multirow{2}{*}{\begin{tabular}[c]{@{}c@{}}\# of Matched \\ \apiseq{} \end{tabular}} & \multirow{2}{*}{\begin{tabular}[c]{@{}c@{}}\# of Matched \\ pairs\end{tabular}} \\
                  &                                                                                    &                                                                                                  &                                                                                 \\ \hline
Python Dataset    & 93                                                                                 & 10267                                                                                            & 1                                                                               \\ \hline
Java Dataset      & 2                                                                                  & 5948                                                                                             & 0                                                                               \\ \hline
\end{tabular}

\end{table}                                 


\subsection{CodeBERT Pretraining (RQ3.1)}
To eliminate the impact of pretraining, we further trained the CodeBERT model from scratch using the task of Python API sequence learning. The model performance is shown in Figure~\ref{codebert-graph_resetparams}. The maximum BLEU-4 score was 35.52\%.  There were fewer iterations in this experiment (40,000 instead of 50,000), but the same experiment with pretraining had already achieved a BLEU-4 score of 40.67\% by 40,000 iterations.   While the result of training CodeBERT from scratch was not as good as the pretrained CodeBERT model, it is still considerably better than the RNN based model, demonstrating that the transformer based model is more effective for encoding the natural language queries and connecting them with the relevant API calls.

\section{Discussion}\label{discussion}

\textit{Model Performance Comparison.} Our study reveals that the transformer based model, in particular CodeBERT, even without self-supervised pretraining,  markedly outperforms the RNN based en-coder-decoder architecture for the task of API sequence learning. When CodeBERT model was pretrained, the convergence is much faster than RNN based models and it further improved the performance evaluated using the BLEU-4 metric. However, such improvement comes with a cost. The CodeBERT model is 674 MB in size when saved to disk, significantly larger than the RNN based model which is only 160 MB. The pretraining process can also be time and resource consuming.

In Table~\ref{table:python-ex} and Table~\ref{table:java-ex}, given in Appendix~\ref{examples}, we present example input and output by different methods from both Python and Java testing sets. We observe that both models still answer many queries incorrectly, revealing a gap before those models can be used in practice. The RNN model sometimes gives the preferable answer, such as its answer to ``round a decimal value'' in Table~\ref{table:java-ex}, while the CodeBERT answer is preferable for other queries, such as ``JSON serialize an object'' in Table~\ref{table:python-ex}.

\textit{Reproducibility.} During our replication of the work by \citet{gu}, we haven't encountered any major challenge thanks to the remarkable effort of releasing and maintaining the source code and dataset on GitHub from the original authors. However, there were a few key differences between our experiment and theirs reported in the paper which might explain the gap of performance on the Java dataset. As we discussed in Section~\ref{subsec:eval_deepapi}, there are a large number of duplicated pairs in the original Java dataset. After removing the duplication, the size shrinks from 7,475,850 to 1,880,472. While the evaluation result using the original dataset can reflect the model performance on common and repeated queries, it might not accurately reveal the model performance on entirely new queries.

Moreover, we adapted the PyTorch implementation of Deep API that was developed by \citet{gu} after their paper was published.  This implementation has significant differences from the original version: for example, the new PyTorch implementation does not weight API calls by their importance, as described in Section 3.2 of the Deep API paper. When we evaluated the performance using the original Theano implementation, the performance is 54.71\% on the original Java dataset which is consistent with the paper, and 20.97\% on the dataset with duplicated instances removed. These results suggest that the significant impact of the dataset quality remains.

Additionally, we did not train the model for as many iterations.  They trained for 1 million iterations (i.e., batches of 100 pairs), while we only trained for 250,000 iterations; but this is unlikely to account for the difference, because our BLEU score had already converged by 250,000 iterations, as shown by Figure~\ref{deepapi-graph_python}.
 
\textit{Limitations and Future Directions.} Table \ref{table:python-ex} in Appendix \ref{examples} gives example inputs and outputs of the models trained on the Python dataset.  From this table we can identify three types of errors. We discuss each of them using representative examples from the table below:
\begin{enumerate}
    \item Some answers are incorrect even though the correct API calls exist in the training set.  For example, the query ``create socket'' has incorrect answers even though the correct API call, \texttt{socket.socket}, appears in the training dataset.  This error might be fixed by tuning the hyperparameters of the models, or by collecting a larger training corpus and training the models for more iterations.
    \item Some answers are incorrect because the correct API calls were removed from the training set to satisfy the upper limit of 10,000 possible output tokens.  For example, the query ``generate md5 hash code'' has incorrect answers because the correct answer, \texttt{hashlib.md5}, was not included in the training set, even though it was found in the original corpus of source code.  To fix this error, a model with a larger maximum vocabulary would be required.
    \item Other answers are incorrect because the correct answer is to call a method on an instance of a class, and method calls are not detected by the dataset curation procedure described in \ref{importstatements}.  For example, the correct answer to the query ``save an image to a file'' is to call the \texttt{save} method of the \texttt{ImageFile} class.  To fix this error, the data curation procedure would have to deduce the types of variables, at least in some cases, so that a call of the form \texttt{image\_file.save} might be recognized as a call to the \texttt{ImageFile} class.
\end{enumerate}

The general method of API discovery pursued by these experiments requires a substantial body of open source software already written in a particular programming language, and it can only learn to recognize APIs that have a history of use in real world software. Ideally, API discovery should work for new languages and new APIs for which there is no example code.  Future experiments might augment the training set with pairs that are extracted from the authoritative API documentation, or the docstrings of the API functions themselves.  This would present some difficulties that would need to be solved. For example, the phrasing of the authoritative API documentation may not reflect how programmers use documentation in their code.  Furthermore, these pairs would account for only a tiny fraction of the training set, so weighting the importance of the training instance depending on its source might be necessary. Additionally, if any of these pairs are assigned to the testing set instead of the training set, then the API call will not appear in the training set at all.  To mitigate these issues, and to avoid overfitting the model with one precise phrasing of the function descriptions, we can apply data augmentation techniques~\cite{nlpda} to generate multiple distinct but equivalent descriptions for each pair that is derived from the authoritative API documentation.

\section{Conclusion}

In this work, we have trained and compared two neural network models to take an input of natural language description of a task and to output a suggested sequence of Python API calls that can perform that task. Previously \citet{gu} had already demonstrated that this could be done with the Java Platform API; we showed that these results can be generalized to at least one other platform Python.  We have repeated our experiment with two different architectures: the RNN Encoder-Decoder architecture achieved a BLEU-4 score of 14.14\% on Python and 12.41\% on Java,  and the CodeBERT architecture achieved a BLEU-4 score of 41.56\% on Python and 24.17\% on Java. We discussed the differences between these two experiments that could account for the improved score of the CodeBERT architecture. Our work calls for future research in combining learning from the wild and the authoritative API documentations.

\begin{acks}
We acknowledge the support of the Natural Sciences and Engineering Research Council of Canada (NSERC)  Discovery Grant Program [RGPIN-2019-05403].
\end{acks}

\bibliographystyle{ACM-Reference-Format}
\bibliography{refs}

\appendix

\newpage
\onecolumn
\section{Example API sequences produced by the models\label{examples}}

\setlength{\twidth}{\dimexpr(.75\paperwidth+4\tabcolsep)\relax}
\begin{table*}[hbt!]
\caption{Example outputs from the RNN (deep API) model and the CodeBERT model trained with Python API calls.  Sequences that accomplish the task are marked with \good, ones that don't are marked with \bad.  If both are \bad, we provide an API sequence that would satisfy the query.  A call of the form X::y means that the method y is called on an instance of the class X.  The queries in this table were adapted from \citet{gu}.\label{table:python-ex}}
\begin{tabular*}{\twidth}{|p{.2\paperwidth}|p{.55\paperwidth}|}
\hline
Query & Python API Sequence \\ \hline

\multirow{2}{\linewidth}{get current time}&\goodrnn time.time\\&\goodcb time.localtime, time.strftime\\ \hline

\multirow{3}{\linewidth}{parse datetime from string}&\badrnn re.match\\&\goodcb pandas.to\_datetime\\ \hline

\multirow{2}{\linewidth}{test file exists}&\goodrnn os.path.join, os.path.exists\\&\goodcb os.path.exists\\ \hline

\multirow{3}{\linewidth}{list files in folder}&\badrnn os.listdir, re.search, os.path.join, os.path.join, os.listdir\\&\goodcb os.walk, os.path.join\\ \hline

\multirow{2}{\linewidth}{match regular expressions}&\badrnn re.escape, re.escape\\&\goodcb re.search\\ \hline

\multirow{3}{\linewidth}{generate md5 hash code}&\badrnn os.path.dirname, os.path.dirname, os.path.abspath, os.path.join, os.path.join, os.path.join\\&\badcb os.urandom\\&\actual hashlib.md5, hashlib.HASH::update, hashlib.HASH::digest\\ \hline

\multirow{3}{\linewidth}{generate random number}&\badrnn numpy.where, scipy.stats.randint, numpy.random.randint\\&\goodcb os.urandom\\ \hline

\multirow{3}{\linewidth}{round a decimal value}&\goodrnn numpy.round\\&\goodcb numpy.round\\ \hline


\multirow{3}{\linewidth}{connect to database}&\badrnn sys.exit\\&\badcb sys.exit\\&\actual mysql.connector.connect\\ \hline


\multirow{3}{\linewidth}{copy file}&\badrnn os.access, time.sleep\\&\badcb os.path.join\\&\actual shutil.copyfile \\ \hline

\multirow{3}{\linewidth}{copy a file and save it to your destination path}&\badrnn os.path.basename\\&\badcb os.path.join\\&\actual shutil.copy\\ \hline

\multirow{2}{\linewidth}{delete files and folders in a directory}&\badrnn os.walk, os.path.join, os.path.dirname, os.path.join, os.rmdir, os.path.join, os.testing, os.rmdir\\&\goodcb os.walk, os.path.join, os.remove, os.rmdir \\ \hline

\multirow{3}{\linewidth}{create socket}&\badrnn time.time, time.time, time.time\\&\badcb time.sleep\\&\actual socket.socket \\ \hline

\multirow{2}{\linewidth}{rename a file}&\goodrnn os.rename\\&\goodcb os.rename\\ \hline

\multirow{3}{\linewidth}{download a file from a url}&\badrnn os.path.expanduser, os.path.basename, os.path.join, os.path.exists, os.path.isfile, os.unlink, os.path.isfile, os.stat, os.path.isdir, os.path.isfile\\&\badcb os.path.exists\\&\actual urllib.request.urlopen, http.client.HTTPResponse::read\\ \hline

\multirow{2}{\linewidth}{JSON serialize an object}&\badrnn sys.stdout.writelines, sys.stdout.flush, sys.stdout.flush, sys.stdout.flush, sys.stdout.flush, sys.stdout.write, sys.stdout.flush\\&\goodcb ujson.dumps\\ \hline


\multirow{2}{\linewidth}{read a binary file}&\goodrnn numpy.memmap\\&\goodcb numpy.fromfile\\ \hline

\multirow{3}{\linewidth}{save an image to a file}&\badrnn imread.imsave.toimage, os.path.dirname, os.path.exists, os.path.dirname, os.makedirs\\&\badcb os.path.splitext\\&\actual  PIL.ImageFile.ImageFile::save\\ \hline


\end{tabular*}

\end{table*}
\setlength{\twidth}{\dimexpr(.75\paperwidth+4\tabcolsep)\relax}
\begin{table*}[h]
\caption{Example outputs from the RNN (deep API) model and the CodeBERT model trained with Java API calls.  Sequences that accomplish the task are marked with \good, ones that don't are marked with \bad.  If both are \bad, we provide an API sequence that would satisfy the query.  The queries in this table were adapted from \citet{gu}.\label{table:java-ex}}
\begin{tabular*}{\twidth}{|p{.2\paperwidth}|p{.55\paperwidth}|}
\hline
Query & Java API Sequence \\ \hline

\multirow{3}{\linewidth}{get current time}&\badrnn new SimpleDateFormat, SimpleDateFormat.parse\\&\badcb new SimpleDateFormat, SimpleDateFormat.format\\&\actual new Date, new SimpleDateFormat, SimpleDateFormat.format \\ \hline

\multirow{2}{\linewidth}{parse datetime from string}&\badrnn String.indexOf, String.length, String.substring, String.length, String.substring\\&\goodcb new SimpleDateFormat, SimpleDateFormat.parse\\ \hline

\multirow{2}{\linewidth}{test file exists}&\badrnn new File, File.isFile, File.getAbsolutePath\\&\goodcb new File, File.exists\\ \hline

\multirow{3}{\linewidth}{list files in folder}&\goodrnn new File, File.listFiles, new ArrayList<File>, File.getName, List<File>.add\\&\badcb new File, File.listFiles\\ \hline

\multirow{2}{\linewidth}{match regular expressions}&\goodrnn Pattern.compile, Pattern.matcher, Matcher.matches\\&\badcb Pattern.compile, Pattern.matcher\\ \hline

\multirow{2}{\linewidth}{generate md5 hash code}&\goodrnn MessageDigest.getInstance, String.getBytes, MessageDigest.digest, Integer.toHexString\\&\goodcb MessageDigest.getInstance, String.getBytes, MessageDigest.digest\\ \hline

\multirow{2}{\linewidth}{round a decimal value}&\goodrnn Math.round, Math.round\\&\badcb Math.pow, Math.round\\ \hline

\multirow{2}{\linewidth}{execute sql statement}&\goodrnn Connection.prepareStatement, PreparedStatement.execute\\&\goodcb Connection.createStatement, Statement.execute, Statement.close\\ \hline

\multirow{2}{\linewidth}{connect to database}&\goodrnn Class.forName, DriverManager.getConnection\\&\goodcb Class.forName, DriverManager.getConnection\\ \hline

\multirow{2}{\linewidth}{create file}&\goodrnn new FileOutputStream, FileOutputStream.write, FileOutputStream.close\\&\goodcb new File, File.createNewFile\\ \hline


\multirow{2}{\linewidth}{copy a file and save it to your destination path}&\goodrnn new File, new FileInputStream, FileInputStream.getChannel, FileOutputStream.getChannel, FileChannel.size, FileChannel.transferTo, FileChannel.close\\&\goodcb new FileInputStream, new FileOutputStream, InputStream.read, OutputStream.write, InputStream.close, OutputStream.close\\ \hline

\multirow{2}{\linewidth}{delete files and folders in a directory}&\goodrnn new File, File.listFiles, File.delete\\&\badcb File.listFiles, File.delete\\ \hline

\multirow{2}{\linewidth}{create socket}&\goodrnn new Socket, Socket.setSoTimeout\\&\badcb Socket.getOutputStream, new ObjectOutputStream\\ \hline

\multirow{2}{\linewidth}{rename a file}&\goodrnn new File, File.renameTo\\&\goodcb new File, File.renameTo\\ \hline

\multirow{2}{\linewidth}{download a file from a url}&\goodrnn new URL, URL.openConnection, URLConnection.getInputStream, new BufferedInputStream, new FileOutputStream, new BufferedOutputStream, BufferedInputStream.read, BufferedOutputStream.write, BufferedInputStream.close, BufferedOutputStream.close\\&\badcb new URL, URL.openConnection, HttpURLConnection.cast, HttpURLConnection.setRequestMethod, HttpURLConnection.setDoInput, HttpURLConnection.setDoOutput, HttpURLConnection.setUseCaches, HttpURLConnection.setInstanceFollowRedirects, HttpURLConnection.setRequestProperty, HttpURLConnection.getOutputStream, OutputStream.write, OutputStream.close, HttpURLConnection.disconnect\\ \hline

\multirow{2}{\linewidth}{XML serialize an object}&\goodrnn new XMLSerializer, new XMLSerializer, XMLSerializer.serialize\\&\badcb new ByteArrayOutputStream, ByteArrayOutputStream.toString\\ \hline

\multirow{3}{\linewidth}{read a binary file}&\badrnn new FileInputStream, new BufferedInputStream, BufferedInputStream.close\\&\badcb new FileInputStream, FileInputStream.close\\&\actual new File, File.length, new FileInputStream, FileInputStream.read, FileInputStream.close\\ \hline

\multirow{2}{\linewidth}{save an image to a file}&\goodrnn new File, ImageIO.write\\&\goodcb new File, ImageIO.write\\ \hline

\multirow{2}{\linewidth}{write an image to a file}&\goodrnn new File, ImageIO.write\\&\goodcb new File, ImageIO.write\\ \hline

\end{tabular*}

\end{table*}

\end{document}